\documentclass[11pt]{article}
\usepackage{a4wide,amssymb,latexsym,epsfig,amsmath,amsfonts,oldgerm,array,graphicx}
\usepackage[english]{babel}
\newcommand{\be}{\begin{equation}}
\newcommand{\ee}{\end{equation}}
\newcommand{\ba}{\begin{array}}
\newcommand{\ea}{\end{array}}
\newcommand{\bea}{\begin{eqnarray}}
\newcommand{\eea}{\end{eqnarray}}
\newcommand{\al}{\alpha}
\newcommand{\bet}{\beta}
\newcommand{\ga}{\gamma}

\newcommand{\D}{\Delta}

\newcommand{\La}{\Lambda}
\newcommand{\cL}{{\cal L}}

\newcommand{\cF}{{\cal F}}
\newcommand{\cB}{{\cal B}}
\newcommand{\cS}{{\cal S}}

\newcommand{\laii}{\lambda_{\rm\scriptscriptstyle I\kern-.1em I}}

\newcommand{\ph}{\phi}

\newcommand{\pa}{\partial}
\newcommand{\R}{{\rm I\kern-0.2em R}}
\newcommand{\C}{{\rm C\kern-0.6em I}}
\newcommand{\N}{{\rm I\kern-0.2em N}}
\newcommand{\Z}{{\sf Z\kern-0.4em Z}}

\newcommand{\ti}{\times}

\newcommand{\na}{\nabla}

\newcommand{\ri}{\right}
\newcommand{\ra}{\rightarrow}

\newcommand{\xx}{{\bf x}}

\newcommand{\vv}{{\bf v}}

\newcommand{\rr}{{\bf r}}
\newcommand{\rrh}{{\hat{\bf r}}}
\newcommand{\BB}{{\bf B}}
\newcommand{\BBh}{\widehat{\bf B}}

\newcommand{\eee}{{\bf e}}

\newcommand{\Tt}{\widetilde{T}}

\newcommand{\St}{\widetilde{S}}

\newcommand{\Bh}{\widehat{B_1}}

\newcommand{\sh}{\hat{s}}

\newcommand{\shn}{\hat{s}_n}

\newcommand{\rd}{{\rm d}}

\newcommand{\cV}{{\cal V}}

\newcommand{\kiiq}{\overline{\kappa}_{\rm\scriptscriptstyle I\kern-.1em I}}
\newcommand{\kiiiq}{\overline{\kappa}_{\rm\scriptscriptstyle
I\kern-.1em I\kern-.1em I}}

\newcommand{\kii}{\kappa_{\rm\scriptscriptstyle I\kern-.1em I}}

\newcommand{\II}{{I\kern-.1em I}}
\newcommand{\III}{{I\kern-.1em I\kern-.1em I}}

\newcommand{\ii}{{\rm\scriptscriptstyle I\kern-.1em I}}
\newcommand{\iii}{{\rm\scriptscriptstyle I\kern-.1em I\kern-.1em I}}

\newcommand{\noi}{\noindent}

\newcommand{\ds}{\displaystyle}
\begin{document}
\title{Radially weighted Backus- and Childress-type bounds for spherical dynamos}
\author{Ralf Kaiser${}^+$ and Andreas Tilgner${}^\#$\\[2ex]
         \small ${}^+$ Fakult\"at f\"ur Mathematik, Physik und Informatik, Universit\"at Bayreuth, D-95440 Bayreuth, Germany\\[1ex]
     \small ${}^\#$ Institut f\"ur Astrophysik und Geophysik, Universit\"at G\"ottingen, D-37077 G\"ottingen, Germany\\[1ex]
     \small ralf.kaiser@uni-bayreuth.de, andreas.tilgner@geo.physik.uni-goettingen.de} 
\date{October, 2025}
\maketitle
\begin{abstract}
In MHD dynamo theory well-known necessary criteria for dynamo action are formulated in terms of lower bounds either on the maximum modulus of the velocity field (Childress-type) or the maximum strain of the velocity field (Backus-type). We generalize these criteria for spherical dynamos by introducing a radially varying weight $f(r)$.
The corresponding {\em l}ower {\em b}ound Reynolds numbers $R_{lb}^C [f]$ (based on velocity) and $R_{lb}^B [f]$ (based on strain) are determined for two types of such weights: a power law profile $f(r) = r^\al$, $0\leq \al \leq 2$ and an optimal radial profile $f_\vv$ depending on the velocity field $\vv$ in question.

To assess the quality of these lower bounds we compare them with weighted critical Reynolds numbers $R_c^C$ (Childress-type) and $R_c^B$ (Backus-type), respectively, for the onset of dynamo action of the well known efficient $s_2t_2$ velocity field (Dudley \& James 1989) and a recently determined ``most efficient'' velocity field (Chen et al.\ 2018). For the latter field we find a Backus-type ratio $R^B_c /R^B_{lb}$ of about $6.4$ with the optimal profile compared to a ratio of about $16.3$ without weight.

\vspace{5mm}

\noi Key Words: Dynamo theory, Spherical dynamos, Backus bound, Variational method.
\end{abstract}
\section{Introduction}
Working dynamos as observed in laboratory experiments and assumed  to operate in stellar or planetary interiors need a sufficiently vigorous velocity field $\vv$ that feeds enough energy into the magnetic field $\BB$ to overcome Ohmic dissipation due to the magnetic diffusivity $\eta$. This process is decribed by the induction equation, an evolution equation for the magnetic field that takes the form 
\be \label{1.1}
\pa_t \BB = \na\ti\big(\vv \ti \BB - \eta \na\ti\BB\big)\, ,\quad \na\cdot\BB = 0 
\ee
or
\be \label{1.1a}
\pa_t \BB + \vv \cdot \na\; \BB = \BB \cdot \na \, \vv - \na \ti (\eta \na \ti \BB)\, , \quad \na \cdot \BB = 0\, , \quad \na \cdot \vv = 0
\ee
in some fluid volume $V$ together with some boundary conditions on $\BB$ and $\vv$ at $\pa V$ (cf.\ Moffatt 1978).

The strength of the driving force is traditionally measured by the space-time maximum value of either $\vv$ itself or -- in view of (\ref{1.1a}) -- of the (symmetrized) velocity-gradient-tensor $\na \vv $. Dimensionless versions of these measures are the Reynolds numbers\footnote{$|\cdot|$ denotes the Euclidean norm in $\R^3$ and $(\cdot\, , \cdot)$ the associated scalar product.}
\be \label{1.2}
R^C := \frac{L}{\eta_0} \, \sup_{V \ti\R_+} |\vv| \qquad \mbox{(Childress-type)}
\ee
and 
\be \label{1.3}
R^B := \frac{L^2}{\eta_0} \, \sup_{V \ti\R_+}\, \max_{|\xx| =1} \big|\big({\cal S} x, x\big)\big|\ ,\quad  {\cal S} := \frac{1}{2}\, \big(\na \vv + (\na \vv)^T\big) \qquad \mbox{(Backus-type)} ,
\ee
where $L$ represents a typical length scale of $V$ and $\eta_0 > 0$ denotes a  lower bound on the magnetic diffusivity $\eta$.
Using, moreover, $L^2/\eta_0$ as time scale, a dimensionless form of the induction equation (\ref{1.1})$_a$ reads for both cases
\be \label{1.4}
\pa_t \BB = R^{B/C}\, \na\ti(\vv^* \ti \BB) - \na \ti(\eta^* \na\ti\BB) 
\ee
with $\vv^*\!\! :=\vv/\sup |\vv|$ resp.\ $\vv /(L \sup \max |({\cal S} x, x)|)$, $\eta^*\! := \eta /\eta_0$, and with
$R^{B/C}$ controlling the relative strength of the driving term over the diffusion term. For given velocity $\vv^*$ and diffusivity $\eta^*$, $R^{B/C}_{c} = R^{B/C}_{c} [\vv^*, \eta^*]$ denote the critical, i.e.\ minimal, values of $R^{B/C}$ that admit non-decaying solutions of eq.\ (\ref{1.4}).\footnote{Throughout this paper we assume $\eta = \eta_0 = const$ so that this dependence no longer needs to be indicated.} 

In view of geo- and astrophysical applications an interesting and intensively investigated class of dynamo models are spherical dynamos with vacuum boundary condition, i.e.\ the magnetic field matches at the bounding sphere continuously to a harmonic field that vanishes at spatial infinity. The classical bounds $R_{lb}^{C} = \pi$ (Childress 1969) and $R_{lb}^{B} = \pi^2$ (Backus 1958) apply to this situation; they have later been improved by Proctor (1977) to $3.50$ and $12.29$, respectively. On the other side early kinematic investigations with simple solenoidal velocity fields and constant diffusivity $\eta^* \equiv 1$ obtained critical Reynolds numbers $R_c^{C}$ of at least about $50$ and often much more (see Dudley \& James 1989 and references therein). The first systematic searches for {\em efficient} velocity fields, i.e.\ with $R_c$ as low as possible, varied only a few flow parameters and found critical Backus-type values of at least about $110$ (Love \& Gubbins 1996, Holme 2003). Only recently this search has been extended to high-dimensional parameter spaces representing in principle the space of all solenoidal steady velocity fields (Chen et al.\ 2018). In terms of Reynolds numbers based on enstrophy 
the authors found a minimum critical Reynolds number a $100$ times larger than a classic lower bound (Proctor 1979) and still $20$ times above a recently improved lower bound (Luo et al.\ 2020). In conclusion the ratio $R_c /R_{lb}$ of most efficient dynamos over best available bounds is still of order $10^1$ in whatever norm. This unpleasant situation can have different causes: the most efficient dynamo has still been missed, bad lower bounds, or unsuitable norms.

In simpler dynamo models this ratio can be much smaller. In the spherically symmetric $\al^2$-mean-field dynamo the ratio can even be lowered to one in a suitable norm (Kaiser \& Tilgner 2018). The velocity field is here replaced by a scalar field with the effect that any $\al$-profile is capable of dynamo action. This is in contrast to the dynamo process as described by eq.\ (\ref{1.4}), where whole classes of velocity fields are excluded from dynamo action by antidynamo theorems, and even if no such theorem applies a velocity field can fail as a dynamo (Kaiser \& Tilgner 1999). So, in this respect the $\al^2$-dynamo with its optimal $R_c/R_{lb}$ ratio may be overly simple.

The introduction of a radial weight is inspired by the {\em m}ost {\em e}fficient flow $\vv_{me}$ by Chen et al.\ (2018), which exhibits a noticeable concentration of flow quantities as well as of the associated magnetic field at the center of the fluid volume. A radially varying weight fits to this situation and, in particular, a power law profile provides a simple means of balancing between center and periphery in the fluid volume. On the other hand, a merely radially varying bound allows the Backus-type variational problem that determines the lower bound to be reduced to a one-dimensional problem that can be solved quite reliably.

All the bounds presented in this note are derived from the fundamental energy balance of either eq.\ (\ref{1.1}) by estimating in the max-norm the velocity or of eq.\ (\ref{1.1a}) by estimating its maximum strain multiplied by the weight, leaving us with a quadratic variational problem for the magnetic field together with the complementary weight (see section 2). Following the unweighted case the use of the poloidal-toroidal decomposition for the magnetic field reduces the vectorial problem to two scalar ones, and expanding these scalars into spherical harmonics leads to a further reduction into one-dimensional problems. The associated Euler-Lagrange equations constitute eigenvalue problems formulated in terms of (systems of) ordinary differential equations  in the radial variable. The toroidal problem is amenable to standard solution techniques and allows even analytical solutions in case of the power law profile. The poloidal problem, however, requires a numerical procedure that can reliably handle the singular terms associated to the power law profile (see section 3). The results expressed in terms of Backus-type ratios $R_c^B/R_{lb}^B$ and Childress-type ratios $R_c^C/R_{lb}^C$ are presented in section 4, whereas conclusions and an outlook on further work are given in section 5. Two more technical questions related to the variational problem determining the lower bound are answered in appendices A and B. 
\section{Energy balance and lower bounds}
In the following we assume the fluid volume $V$ to be a ball $B_1$ of radius $L =1$ with boundary $S_1$ and exterior (vacuum) region $\Bh$. The magnetic field $\BB$, the velocity field $\vv$, and the diffusivity $\eta$ are variable in the spatial coordinate $\rr \in B_1$ and time $t \in \R_+$. The diffusivity is positively bounded from below, $\eta \geq \eta_0 > 0$, and the velocity field is bounded in the sense of $R^C < \infty$ (Childress-case) or $R^B < \infty$ (Backus-case). Moreover, in the Backus-case the velocity field is assumed to be solenoidal. Furthermore, at the boundary $S_1$, the velocity field has a vanishing radial component and the magnetic field matches continuously to some harmonic field that vanishes at infinity. 

Solutions $\BB$ of the induction equation (\ref{1.1}) satisfy an energy balance, which is obtained by scalar multiplication of eq.\ (\ref{1.1})$_a$ with $\BB$ and integration over $B_1$. By proper reformulation of the boundary terms one obtains (Backus 1958):
$$
\frac{1}{2} \frac{\rd}{\rd t}
\int_{\R^3} |\BB |^2\, \rd v  = \int_{B_1} (\vv \ti \BB) \cdot (\na \ti \BB)\, \rd v - \int_{B_1} \eta\, |\na \ti \BB|^2\, \rd v \, .
$$
Suitably estimating the first term on the right-hand side yields the classical Childress criterion. We repeat this estimate, however, enlarged by the neutral expression $g \cdot g^{-1}$, where $g$ denotes a radial weight $g: (0,1) \ra \R_+$:
\be \label{2.1}
\ba{l}
\ds \frac{1}{2}\, \frac{\rd}{\rd t} \int_{\R^3} |\BB(\cdot, t)|\, \rd v \leq \int_{B_1} |\vv(\cdot ,t)\, g|\, |\BB(\cdot ,t)\, g^{-1}|\,|\na \ti \BB(\cdot ,t)|\, \rd v 
- \eta_0 \int_{B_1} |\na \ti \BB(\cdot ,t)|^2\, \rd v \\[1em]
\qquad \ds \leq \sup_{t \in \R_+} \, \sup_{\rr \in B_1}\Big\{ |\vv(\rr, t)|\, g(|\rr|)\Big\}\bigg(\int_{B_1} |\BB(\cdot, t)|^2\, g^{-2}\, \rd v\bigg)^{1/2}
\bigg(\int_{B_1} |\na \times \BB(\cdot, t)|^2\, \rd v\bigg)^{1/2} \\[1em]
\qquad \qquad \qquad \qquad \ds - \eta_0 \int_{B_1} |\na \ti \BB(\cdot ,t)|^2\, \rd v \\[1em]
\qquad \qquad \ds  = \eta_0\,\Bigg[ R^C \Bigg(\frac{\int_{B_1} |\na \times \BB(\cdot, t)|^2\, \rd v}{\int_{B_1} |\BB(\cdot, t)|^2\, g^{-2}\, \rd v}\Bigg)^{-1/2} - 1 \Bigg] \int_{B_1} |\na \times \BB(\cdot, t)|^2 \,\rd v
\ea
\ee
with the weighted Childress-type Reynolds number 
\be \label{2.2}
R^C := R^C [\vv, g] :=  \frac{1}{\eta_0}\, 
 \sup_{t \in \R_+} \, \sup_{\rr \in B_1}\Big\{ |\vv(\rr, t)|\, g(|\rr|)\Big\}\, .
\ee
Inequality (\ref{2.1}) suggests to solve the variational problem 
\be \label{2.3}
\inf_{0 \neq \BB \in \cB}\, \frac{\ds\int_{B_1} |\na \times \BB|^2\, \rd v}{\ds \int_{B_1} |\BB|^2\, g^{-2}\, \rd v} =:\big(R^C_{lb}\, [g^2] \big)^2 =: {R^C_{lb}}^{\, 2} ,
\ee
where\footnote{Readers not familiar with ``weak differentiability'' may ignore the additive ``weak'' without missing the essence of the presentation.}
\be \label{2.3a}
\ba{l}
\ds \cB := \big\{ \BB : B_1 \ra \R^3 :\BB \mbox{ is weakly differentiable, } \na \cdot \BB = 0, \mbox{ and both  integrals in (\ref{2.3})} \\
\qquad\quad  \ds \mbox{ are finite; moreover } \BB  \mbox{ allows a harmonic extension } \BBh: \Bh \ra \R^3, \\
\qquad \quad \mbox{ such that  } \BB =\BBh \mbox{ at } S_1, \mbox{ and } |\BBh| \ra 0 \mbox{ for } |\rr| \ra \infty \, .
\big\}\, .
\ea
\ee
An equivalent formulation of the variational problem (\ref{2.3}) is given by 
\be \label{2.4}
\inf_{0 \neq \BB \in W_0^1}\, \frac{\ds\int_{\R^3} |\na \times \BB|^2\, \rd v}{\ds \int_{B_1} |\BB|^2\, g^{-2}\, \rd v} =\, \big(R^C_{lb}\, [g^2] \big)^2 
\ee
with the simpler variational set
$$
\ba{l}
\ds W_0^1 := \big\{ \BB : \R^3 \ra \R^3 :\BB \mbox{ is weakly differentiable, } \na \cdot \BB = 0,\; |\BBh| \ra 0 \mbox{ for } |\rr| \ra \infty, \\
\qquad \qquad \mbox{ and both integrals in (\ref{2.4}) are finite.} 
\big\}\, .
\ea
$$
In fact, considering the Euler-Lagrange equations associated to the variational problem (\ref{2.4}), the minimizer in (\ref{2.4}) turns out to be harmonic in the exterior region $\Bh$ and thus to be included into the variational set $\cB$. The existence of a positive infimum in the variational problem depends on the weight $g$ (see below); only in the case that $g$ is positively bounded from below a positive minimum is well established.

Once the number $R_{lb}^C$ is determined inequality (\ref{2.1}) takes the final form
\be \label{2.5}
\ba{l}
\ds \frac{\rd}{\rd t} \int_{\R^3} |\BB(\cdot, t)|\, \rd v 
\leq \eta_0\,\Big( R^C [\vv,g] /R_{lb}^C [g]  - 1 \Big) \int_{B_1} |\na \times \BB(\cdot, t)|^2 \,\rd v \\[1em]
\qquad \qquad \qquad \qquad \ds \leq \pi^2\, \eta_0\,\Big( R^C [\vv,g] /R_{lb}^C [g]  - 1 \Big) \int_{\R^3} |\BB(\cdot, t)|^2 \,\rd v\, ,
\ea
\ee
where in the last line we made use of the classical variational inequality (Backus 1958):
$$
\int_{B_1} |\na \ti \BB|^2\, \rd v \geq \pi^2 \int_{\R^3} |\BB|^2\, \rd v\, .
$$
Inequality (\ref{2.5}) implies an exponentially in time decaying bound onto the magnetic energy for any Reynolds number $R^C [\vv, g]$ below $R^C_{lb} [g]$.

The Backus-case differs from the Childress-case by a reformulation of the interaction term in the energy balance using the divergence-constraint on the velocity field. Inequality (\ref{2.1}) then takes the form
\be \label{2.5a}
\ba{l}
\ds \frac{1}{2}\,\frac{\rd}{\rd t} \int_{\R^3} |\BB(\cdot, t)|\, \rd v \leq \int_{B_1} |\BB(\cdot , t)\, \cS (\cdot , t)\, \BB(\cdot , t)|\, \rd v 
- \eta_0 \int_{B_1} |\na \ti \BB(\cdot ,t)|^2\, \rd v \\[1em]
\qquad \ds \leq \sup_{(\rr,t) \in B_1 \ti \R_+}\, \max_{|\xx| =1} \Big\{ \big|\big(\cS(\rr, t) \xx\, , \xx\big)\big|\, f(|\rr|)\Big\}\, \int_{B_1} |\BB(\cdot, t)|^2\, f^{-1}\, \rd v \\[1em]
\qquad \qquad \qquad \qquad \qquad \quad \ds - \eta_0 \int_{B_1} |\na \ti \BB(\cdot ,t)|^2\, \rd v \\[1em]
\qquad \qquad \ds  = \eta_0\,\Bigg[ R^B \Bigg(\frac{\int_{B_1} |\na \times \BB(\cdot, t)|^2\, \rd v}{\int_{B_1} |\BB(\cdot, t)|^2\, f^{-1}\, \rd v}\Bigg)^{-1} - 1 \Bigg] \int_{B_1} |\na \times \BB(\cdot, t)|^2 \,\rd v
\ea
\ee
with weight $f: (0, 1) \ra \R_+$ and weighted Backus-type Reynolds number 
\be \label{2.6}
R^B := R^B [\vv, f] :=  \frac{1}{\eta_0}\, 
\sup_{(\rr,t) \in B_1 \ti \R_+}\, \max_{|\xx| =1} \Big\{ \big|\big(\cS(\rr, t) \xx\, , \xx\big)\big|\, f(|\rr|)\Big\}\, .
\ee
The relevant variational problem is here
\be \label{2.7}
\inf_{0 \neq \BB \in \cB}\, \frac{\ds\int_{B_1} |\na \times \BB|^2\, \rd v}{\ds \int_{B_1} |\BB|^2\, f^{-1}\, \rd v} =: R^B_{lb}\, [f]  =: {R^B_{lb}}\, ,
\ee
and $R^B < R^b_{lb}$ implies again an exponentially decaying bound for the magnetic energy.

According to (\ref{2.3}) and (\ref{2.7}) the lower bounds are obviously related by 
\be \label{2.8}
R^B_{lb} [f] = \big(R_{lb}^C [f]\big)^2 .
\ee
Note however, that according to (\ref{2.2}) and (\ref{2.6}) $R^C$ and $R^B$ measure different flow quantities.
\section{Variational problems with radial weights}
The quadratic form of the variational expression
\be \label{3.0}
\cV\, [\BB] = \frac{\ds\int_{B_1} |\na \times \BB|^2\, \rd v}{\ds \int_{B_1} 
|\BB|^2\, f^{-1}\, \rd v} 
\ee
with purely radial weight $f$ allows the exact splitting of the vectorial problem into two scalar ones by means of the ($L^2 (S_1)$-orthogonal) poloidal-toroidal decomposition of solenoidal vector fields.
The spherical symmetry of the weight allows, moreover, the exact minimization with respect to the angular variables so that we are left with the much simpler but still nontrivial problem of a one-dimensional scalar minimization. This reduction process is quite standard and we restrict the presentation to the minimum.

The magnetic field $\BB$ is represented by the toroidal scalar $T$ and the poloidal one $S$ in the form
\be \label{3.1}
\BB = -\La T -\na \ti \La S\, ,
\ee
where $\La$ denotes the non-radial-derivative operator $\rr \ti \na$. Its square $\La \cdot \La=: \cL$ is the Laplace-Beltrami-operator on the unit sphere $S_1$; $-\cL$ is a positive symmetric operator with the spherical harmonics $Y_{nm}$ as eigenfunctions:
$$
- \cL\,  Y_{nm}  = n(n + 1) Y_{nm}\, ;
$$
moreover, $\cL$ is relatd to the Laplacian $\D$ by
$$
\D = \frac{1}{r}\, \pa_r^2\, r + \frac{1}{r^2}\, \cL\, ,
$$
where $r := |\rr|$, $\rrh:= \rr/r$, and $\pa_r:= \rrh \cdot \na$.

Standard manipulations then yield
$$
\ba{rl}
\ds \int_{B_1} |\BB|^2 \, f^{-1}\, \rd v \!\!\! & = \ds \int_{B_1} |\La\, T + \na \ti \La\, S|^2 \,f^{-1}\, \rd v  = \ds \int_{B_1} \big(|\La\, T|^2 + |\na \ti \La\, S|^2\big) f^{-1}\, \rd v \\[1.5em]
& = \ds \int_{B_1} \Big(|\La\, T|^2 + \Big(\frac{1}{r}\cL\, S\Big)^2 + \Big|\La\, \frac{1}{r} \pa_r (r  S)\Big|^2\Big) f^{-1}\, \rd v
\ea
$$
and
$$
\ba{rl}
\ds \int_{B_1} |\na \ti \BB|^2 \rd v \!\!\! & = \ds \int_{B_1} |\na \ti (\La\, T + \na \ti \La\, S|^2 \,\rd v \\[1.5em]
& = \ds \int_{B_1} \big(|\na \ti \La\, T|^2 + |\na \ti (\na \ti \La\, S) |^2\big) \rd v \\[1.5em]
& = \ds \int_{B_1} \Big(\Big(\frac{1}{r} \cL\, T\Big)^2 + \Big|\La\,\frac{1}{r} \pa_r ( r T)\Big|^2 + |\La\, \D\, S|^2 \Big) \rd v\, .
\ea
$$
By the elementary inequality
\be \label{3.4}
\sum_{n=1}^N a_n\, \Big/ \sum_{n=1}^N b_n \, \geq \, \min_{1 \leq n \leq N} \frac{a_n}{ b_n}\, ,\quad a_n \geq 0\, ,\quad b_n > 0\, ,\quad N \in \N
\ee
the variational expression (\ref{3.0}) can thus be separated into a toroidal and a poloidal part:
\be \label{3.4a}
\ba{l}
\ds \cV\, [\BB] \geq \min \Bigg\{ \frac{\int_{B_1} \Big(\Big(\frac{1}{r} \cL\, T\Big)^2 + \Big|\La\,\frac{1}{r} \pa_r ( r T)\Big|^2 \Big) \rd v}{ \int_{B_1} \Big(|\La\, T|^2 \Big) f^{-1}\, \rd v }\, , \,
\frac{\int_{B_1} |\La\, \D\, S|^2 \, \rd v}{\int_{B_1} \Big( \Big(\frac{1}{r}\cL\, S\Big)^2 + \Big|\La\, \frac{1}{r} \pa_r (r  S)\Big|^2\Big) f^{-1}\, \rd v} \Bigg\} \\[2em]
\qquad \quad \ds =: \min \Big\{ \cV_t [T]\, ,\cV_p [S]\, \Big\}\, .
\ea
\ee
Expanding the variables $r T=:\Tt$ and $r S =:\St$ into spherical harmonics $\big\{ Y_{nm} : n \in \N ,|m| \leq n \big\}$ according to 
$$
\Tt(\rr) = \sum_{n,m} \Tt_{nm} (r)\, Y_{nm}(\rrh) \; , \qquad    \St(\rr) = \sum_{n,m} \St_{nm} (r)\, Y_{nm} (\rrh)
$$
allows by $L^2 (S_1)$-orthogonality  of the $Y_{nm}$ and again with (\ref{3.4}) the further reduction\footnote{By taking suitable linear combinations the $Y_{nm}$ can assumed to be real and to satisfy standard orthogonality relations; the prime means the derivative $\frac{\rd}{\rd r}$.} 
\be \label{3.5}
\ba{l}
\ds \inf_{\Tt} \, \cV_t[\Tt] = \inf_{\{\Tt_{nm}\}} \frac{\ds \sum_{n,m} n(n+1) \int_0^1 \Big(\frac{n(n+1)}{r^2}\, \Tt^2_{nm} + (\Tt'_{nm})^2 \Big) \rd r}{\ds \sum_{n,m} n(n+1) \int_0^1 \Tt_{nm}^2\, f^{-1}\, \rd r} \\[2em]
\ds \qquad \qquad \; \geq \lim_{N \ra \infty}\! \min_{\scriptsize{
\ba{c}
n \leq N  \\ |m| \leq n \ea
}}\! 
\inf_{\Tt_{nm}}\, \frac{\ds \int_0^1 \Big(\frac{n(n+1)}{r^2}\, \Tt^2_{nm} + (\Tt'_{nm})^2 \Big) \rd r}{\ds \int_0^1 \Tt_{nm}^2\, f^{-1}\, \rd r} 
\ea 
\ee
and 
\be \label{3.6}
\ba{l}
\ds \inf_{\St} \, \cV_p[\St] = \inf_{\{\St_{nm}\}} \frac{\ds \sum_{n,m} n(n+1) \int_0^1 \Big(\St''_{nm} - \frac{n(n+1)}{r^2}\, \St_{nm} \Big)^2 \rd r}{\ds \sum_{n,m} n(n+1) \int_0^1 \Big(\frac{n(n+1)}{r^2}\, \St_{nm}^2 + (\St'_{nm})^2\Big) f^{-1}\, \rd r} \\[2em]
\ds \qquad \qquad \; \geq \lim_{N \ra \infty}\! \min_{\scriptsize{
\ba{c}
n \leq N  \\ |m| \leq n \ea
}}\! 
\inf_{\St_{nm}}\, 
\frac{\ds \int_0^1 \Big(\St''_{nm} - \frac{n(n+1)}{r^2}\, \St_{nm} \Big)^2 \rd r}{\ds \int_0^1 \Big(\frac{n(n+1)}{r^2}\, \St_{nm}^2 + (\St'_{nm})^2\Big) f^{-1}\, \rd r} \, .
\ea 
\ee
In (\ref{3.5}) the minimum is clearly reached at $(n,m) = (0,0)$ with the result:
\be \label{3.7}
\inf_{\Tt} \cV_t [\Tt] = \inf_{0 \neq t \in H^1_0} \frac{\ds \int_0^1 \Big({t'}^2 + \frac{2}{r^2}\, t^2\Big) \rd r}{\ds \int_0^1 t^2\, f^{-1} \,\rd r} =: \mu^t [f] =: \mu^t .
\ee
The variational class $H_0^1$ follows with (\ref{3.1}) from (\ref{2.3a}) and is given by 
$$
\ba{l}
\ds H_0^1 := \big\{ t : (0,1) \ra \R : t \mbox{ is weakly differentiable, }  t(0) = t(1) =0\, , \\[.3em]
\qquad \qquad \mbox{ and both integrals in (\ref{3.7}) are finite.} 
\big\}\, .
\ea
$$
In (\ref{3.6}) the minimizing $n$ is not as obvious; so we keep the index $n$ but skip the dummy index $m$ to obtain the poloidal problem 
\be \label{3.8}
\inf_{0 \neq s \in H^2_n} \frac{\ds \int_0^1 \Big(s_n'' - \frac{n(n+1)}{r^2}\, s_n\Big)^2 \rd r}{\ds \int_0^1 \Big({s_n'}^2 + \frac{n(n +1)}{r^2}\, s_n^2 \Big) f^{-1} \,\rd r} =: \mu_n^p [f] =: \mu^p_n
\ee
with variational class
$$
\ba{l}
\ds H_n^2 := \big\{ s_n : (0,1) \ra \R : s_n \mbox{ is twice weakly differentiable, }  s_n (0) = s'_n (0) =0\, , \\[.3em]
\qquad \qquad \quad s_n (1) + n\, s'_n (1) =0\,\mbox{ and both integrals in (\ref{3.8}) are finite.} 
\big\}\, .
\ea
$$
A necessary and sufficient condition on the weight $f$ to obtain  nontrivial lower bounds of type (\ref{2.7}) takes the form 
\be \label{3.9}
|f(r)| \geq m\, r^2 \qquad \mbox{on }\; (0,1)
\ee
for some $m > 0$. The necessity is demonstrated in appendix A, whereas sufficiency follows by Hardy's inequality (see eqs.\ (\ref{B.2}) in appendix B) applied to the variational expressions (\ref{3.7}) and (\ref{3.8}).

For weights of type (\ref{3.9}) we prove in appendix B the inequality 
\be \label{3.10}
\mu_n^p \geq \min \big\{\mu_1^p ,\, 
m \big( n(n+1) - 3/2\big) \big\}\, ,
\ee
which delimits the minimum to those $n$ that satisfy $n (n+1) < \mu_1^p/m + 3/2$ (and, in fact, for the weights under consideration, to $n=1$).

The Euler-Lagrange equation associated to the variational expression (\ref{3.7}) reads
\be \label{3.11}
t'' + \Big( \mu^t f^{-1} - \frac{2}{r^2} \Big) t = 0\quad \mbox{ on }\, (0,1)\, ,\qquad t(0) = t(1) = 0
\ee
with weight $f$ and Lagrange-parameter $\mu^t$. For $f=1$ the solution can be expressed by the spherical Bessel function $j_1$ with the 
well-known result $\mu^t = (i_{1,1})^2 \approx 20.19$, where $i_{1,1}$ denotes the first nontrivial zero of $j_1$ (Backus 1958).

The Euler-Lagrange equation associated to (\ref{3.8}) reads 
\be \label{3.12}
\ba{c}
\ds  D_n D_n s_n + \mu_n^p \big( f^{-1} D_n s_n + (f^{-1})' s_n' \big) = 0 \quad  \mbox{ on }\, (0,1)\, , \\[1em]
\ds s_n (0) =s_n' (0) = 0\, , \quad s_n' (1) + n\, s_n (1) = 0\, ,\quad \big[ (D_n s_n)' + n (D_n s_n - \mu_n^p s_n)\big]_{r =1} = 0
\ea
\ee
with the second-order operator $D_n := \pa_r^2 - \frac{n(n+1)}{r^2}$. In the classical case $f =1$, eq.\ (\ref{3.12}) with $n=1$ separates into
$$
(D_1 + \mu_1^p)\, D_1 s_1 = 0\, ,
$$
which also allows an analytic solution resulting into the poloidal minimum $\mu^p = \mu_1^p = \ga^2 \approx 12.29$, where $\ga$ is the smallest nontrivial solution of the equation $3 j_1 (\ga ) + 2 \ga j_0 (\ga ) = 0$ (Proctor 1977).

For nonconstant weights problems (\ref{3.11}) and (\ref{3.12}) are amenable to standard numerical solution techniques such as the shooting method which is the most basic numerical technique for solving eigenvalue and boundary value problems. However, shooting methods integrate differential equations from
one boundary to the other and are easily compromised by round off error near the starting boundary if that error is amplified by singular terms at that boundary (as is the case here), or by exponentially growing solutions within the fundamental system of the ordinary differential equation.

A more tolerant numerical scheme is obtained if one
discretizes directly the variational problems (\ref{3.7}) and (\ref{3.8}) and thus solves finite-dimensional variational problems instead of singular Euler-Lagrange equations.
To this end, we represent the optimizers in equations (\ref{3.7}) and (\ref{3.8}) as linear combinations of $N$ base functions which are $N$ linear independent combinations of the first $N+M$ Chebyshev polynomials satisfying the $M$ boundary conditions in problem (\ref{3.7}) ($M=2$) or (\ref{3.8}) ($M$=3). These
linear combinations are determined as part of the following numerical procedure: Construct an $M \times (N+M)$ dimensional matrix by imposing the $M$ boundary conditions on the $N+ M$  dimensional coefficient vector representing the first $N+M$
Chebyshev polynomials. A singular value decomposition of that matrix returns a
base of its $N$ dimensional null space which corresponds to the $N$ requested linear combinations. 

Then, collect the $N$ coefficients in this
representation of the optimizing function in a vector $\bf x$ and code matrices such that problems (\ref{3.7}) and (\ref{3.8}) ask for optimizations of expressions of the form
$({\cal M}_1 {\bf x} , \xx) / ({\cal M}_2 {\bf x} , \xx)$
where ${\cal M}_1$ and ${\cal M}_2$ are symmetric matrices. Determine the eigenvalues $\lambda$ of the finite-dimensional
generalized eigenvalue problem 
${\cal M}_1 {\bf x} = \lambda \, 
{\cal M}_2 {\bf x}$.
The smallest of these eigenvalues is the numerical approximation for $\mu^t$ or $\mu_n^p$. The optima obtained this way are constant to within better than 1\% for $16 \leq N \leq 128$ and constitute the results given in the following figures and tables.
\section{Weighted bounds}
In the following we consider two types of weights, viz., a universal weight of power law type $f(r) = r^\al$, $0\leq \al \leq 2$ and a radial profile $f_\vv$ that is in a sense optimally adapted to a given velocity field $\vv$. To assess the quality of the lower bound $R_{lb}$ we consider the ratio $R_c/R_{lb}$, where $R_c$ denotes the (weighted) critical Reynolds number of a given velocity field and compare, in particular, the weighted with the unweighted case.

The main motivation for considering a weight of power law type stems from a recently determined optimal kinematic dynamo (Chen et al.\ 2018), whose velocity field exhibits a remarkable concentration of flow quantities at the center of the fluid volume (see Fig.\ 1 below and Fig.\ 7 in (Chen et al.\ 2018)). 
\begin{figure} \label{Fig1}
\includegraphics[width=0.8\linewidth]{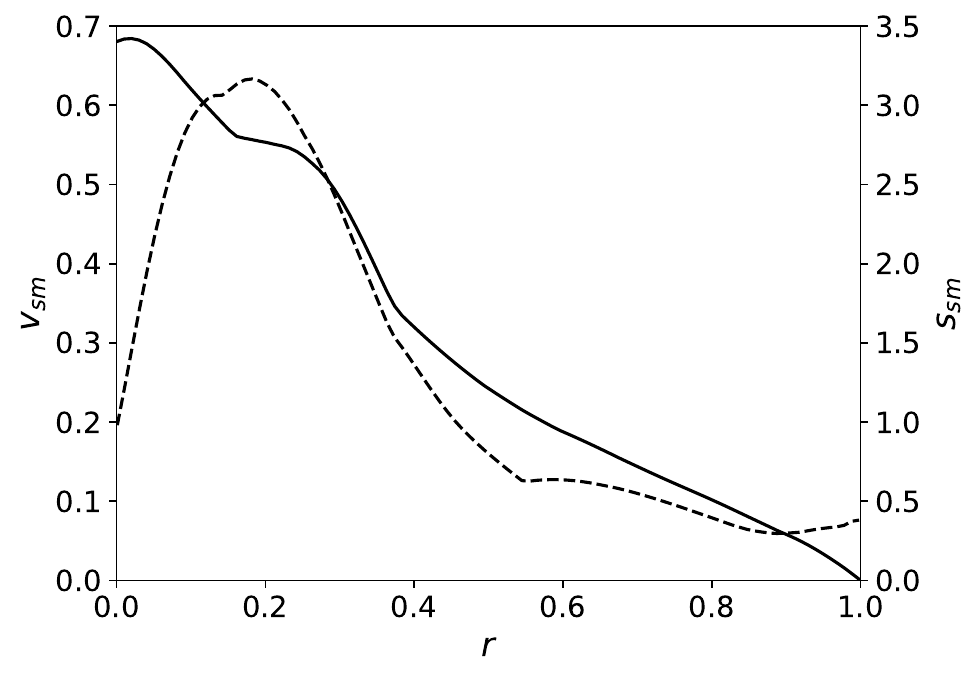}
\caption{Spherically maximized modulus of velocity $v_{sm}$ (solid line) and of strain $S_{sm}$ (dashed line) versus radius $r$ for the most efficient velocity field $\vv_{me}$ of Chen et al.\ (2018) in units given there (cf.\ definitions (\ref{4.3}) and (\ref{4.5})).}
\end{figure}
To determine this dynamo the authors used a variational search method in the space of all steady solenoidal velocity fields, vanishing at the boundary, to obtain the most efficient field\footnote{A representation of this field in terms of spherical harmonics $Y_{nm}$ with $n \leq 24$ and $|m| \leq 24$ and a polynomial basis in radial direction containing powers up to $r^{49}$ can be found in the supplementary material linked to (Chen et al.\ 2018).}
$\vv_{me}$ in the sense that the associated magnetic field has maximum growth rate after some fixed time interval and for fixed Reynolds number $R^{ens}$, measured in the enstrophy norm. At the vanishing point of this maximum growth rate the authors found the critical value $R_c^{ens} = 64.45$, which is in fact lower than for any other known dynamo in this class, and the authors argue to have determined the global minimum. A lower bound in this norm, viz.\ $R_{lb}^{ens} = 3.10$, has recently been given by Luo et al.\ (2020), so that the ratio $R^{ens}_c /R^{ens}_{lb}$ takes the value 20.8. Switching to Childress- or Backus-type Reynolds numbers lowers this ratio:
$$
R_c^C /R_{lb}^C = 44.1/3.5 \approx 12.6\; ,\qquad 
R_c^B /R_{lb}^B = 204/12.3 \approx 16.6\; ,
$$
and further lowering can be expected by introducing radially varying weights.\footnote{Note that the value of $R_c^B$ we computed with the field given by Chen et al.\ (2018) differs by 5\% from the value given in table 4 of this reference.}

Concerning the power law weight $f(r) = r^\al$ a nontrivial lower bound $R^B_{lb} [r^\al] =: R^B_{lb} (\al)$ can only be expected for $\al \leq 2$ as demonstrated in appendix A. The toroidal minimum $\mu^t (\al)$ can be explicitly expressed as
\be \label{4.1}
\mu^t(\al) = \big((1 -\al/2) I_{3/(2-\al), 1} \big)^2 , \qquad \al < 2\, ,
\ee
where $I_{\bet, 1}$ is the first nontrivial zero of the Bessel function $J_\bet$. In fact
$$
t: r \mapsto \sqrt{r}\, J_{3/(2 -\al)} \big(2 \sqrt{\mu^t} /(2-\al)\, r^{1-\al/2} \big)
$$
is the solution regular at the origin of the Euler-Lagrange equation (\ref{3.11}) with $f(r) = r^\al$, and the condition $t(1) = 0$ fixes $\mu^t (\al) $ at the value (\ref{4.1}).

The limit case $\al = 2$ requires a special consideration  (which is omitted here) and results in  $\mu^t (2) = 9/4$.

In the poloidal case an analytic solution of the Euler-Lagrange equation (\ref{3.12}) is not available and the numerical determination of the poloidal minima $\mu_n^p$ starts directly at the variational expression (\ref{3.8}) as explained at the end of the last section. 


\begin{figure} \label{Fig2}
\includegraphics[width=0.7\linewidth]{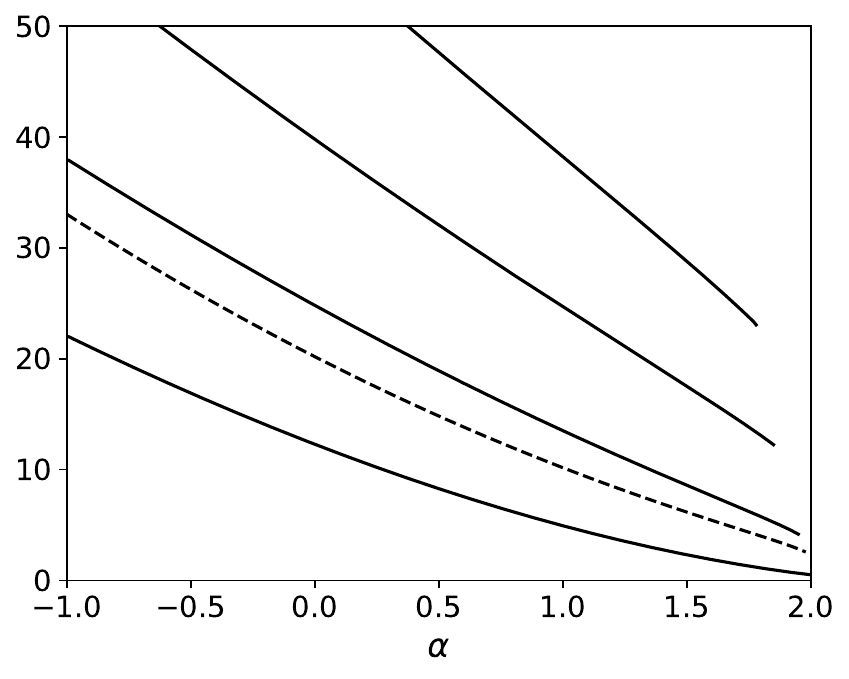}
\caption{Poloidal minima $\mu_n^p$ for $n= 1,\ldots, 4$ (solid lines, increasing with $n$) and toroidal minimum $\mu^t$ (dashed line) versus $\al$ for the power law weight $f(r) = r^\al$.}
\end{figure}
Figure 2  shows the poloidal minima $\mu_n^p (\al)$ for $n= 1,\ldots, 4$ and $\al < 2$ together with the toroidal minimum $\mu^t (\al)$. The global minimum is clearly taken by $\mu_1^p$, which holds by (\ref{3.10}) and $\al  \in [0, 2]$ in fact for all $n \in \N$. By (\ref{2.7}), (\ref{3.0}), (\ref{3.4a})--(\ref{3.8}) we thus obtain for the Backus-type lower bound
$$
R_{lb}^B (\al) = \mu_1^p (\al)\, .
$$
The monotonous decay with respect to $\al$ clearly reflects the monotonous growth of the weight with $\al$  on the interval $(0,1)$. On the other hand  
computing the Backus-type critical Reynolds number $R_c^B [\vv_{me}, r^\al ] = R_c^B (\al)$ for the most efficient flow $\vv_{me}$ according to (\ref{2.6}) and for constant diffusivity $\eta = \eta_0$ one finds also a decreasing function  with respect to $\al$ (see Fig.\ 3) reflecting the fact that the weight increasingly suppresses the flow at the center of the ball. 
\begin{figure}
\includegraphics[width=0.8\linewidth]{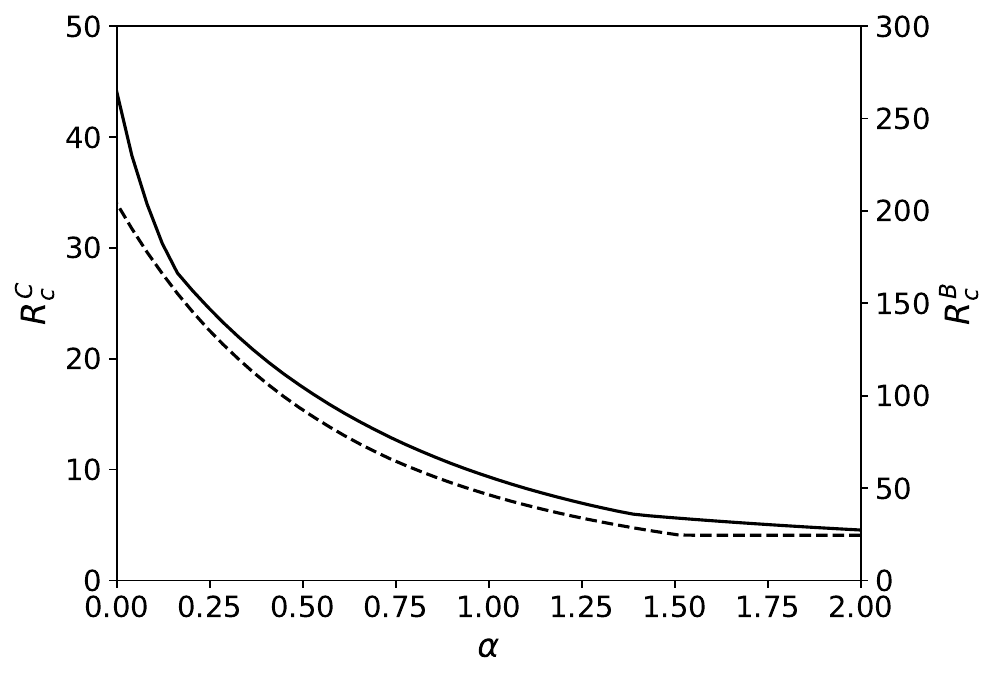}
\caption{Childress-type (solid line) and Backus-type (dashed line) critical Reynolds numbers $R_c^C$ with weight $g= r^{\al/2}$ and $R_c^B$ with weight $f = r^{ \al}$ versus $\al$ for the velocity field $\vv_{me}$ (cf.\ definitions (\ref{2.2}) and (\ref{2.6})).}
\end{figure}
The ratio $R_c^B (\al) /R_{lb}^B (\al)$ balances both effects  and takes in fact  a minimum value of $9.36$ at $\al = 1.07$ (see Fig.\ 4). 
\begin{figure}
\includegraphics[width=0.8\linewidth]{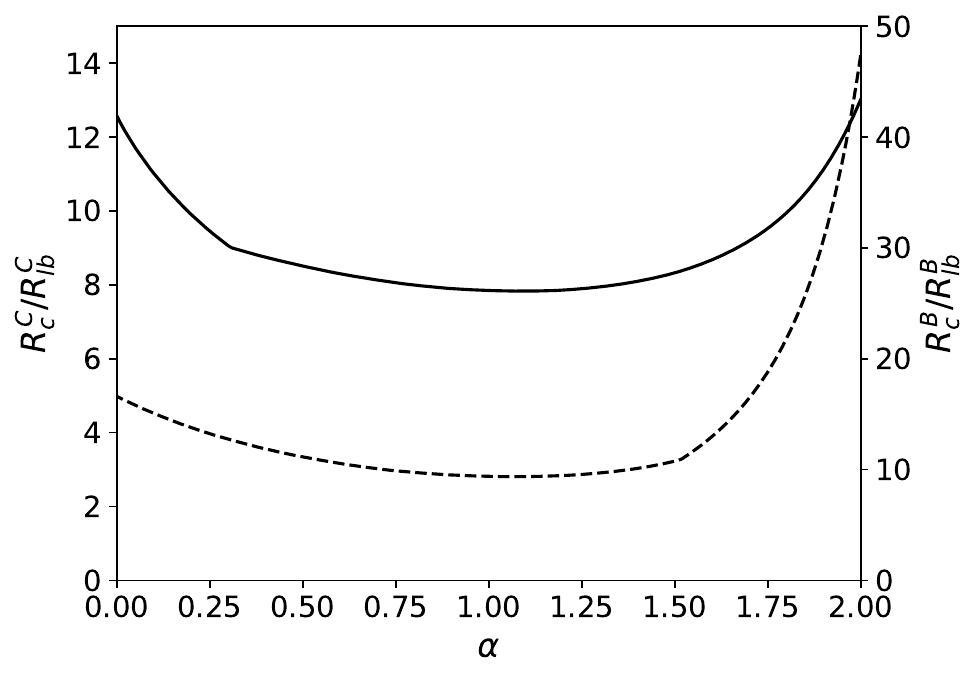}
\caption{Childress-type ratio $R_c^C/R_{lb}^C$ (solid line)  and Backus-type ratio $R_c^B/R_{lb}^B$ (dashed line) with power law weight versus $\al$ for the velocity field $\vv_{me}$.}
\end{figure}

A similar effect is observed in the Childress case. Setting $g^2 = r^\al$, computing by (\ref{2.2}) the Childress-type critical Reynolds number $R_c^C [\vv_{me} , r^{\al/2}] =: R_c^C (\al)$, and observing (\ref{2.8}), we find for the ratio $R_c^C (\al)/R_{lb}^C (\al)$ the even lower minimum of $7.83$ at $\al = 1.08$ (see Fig.\ 4). Note that velocity field and strain do not vanish at the origin. This puts the lower limit $\al =0$ on the admissible $\al$-range to obtain finite numbers $R_c^{B/C} (\al)$.

For comparison we consider another well known efficient velocity field, viz., the $s_2t_2$ field $\vv_{D\! J}$ of Dudley \& James (1989), which, however, does not show a significant concentration of flow quantities at the center of the fluid volume (see Fig.\ 5). 
\begin{figure}
\includegraphics[width=0.8\linewidth]{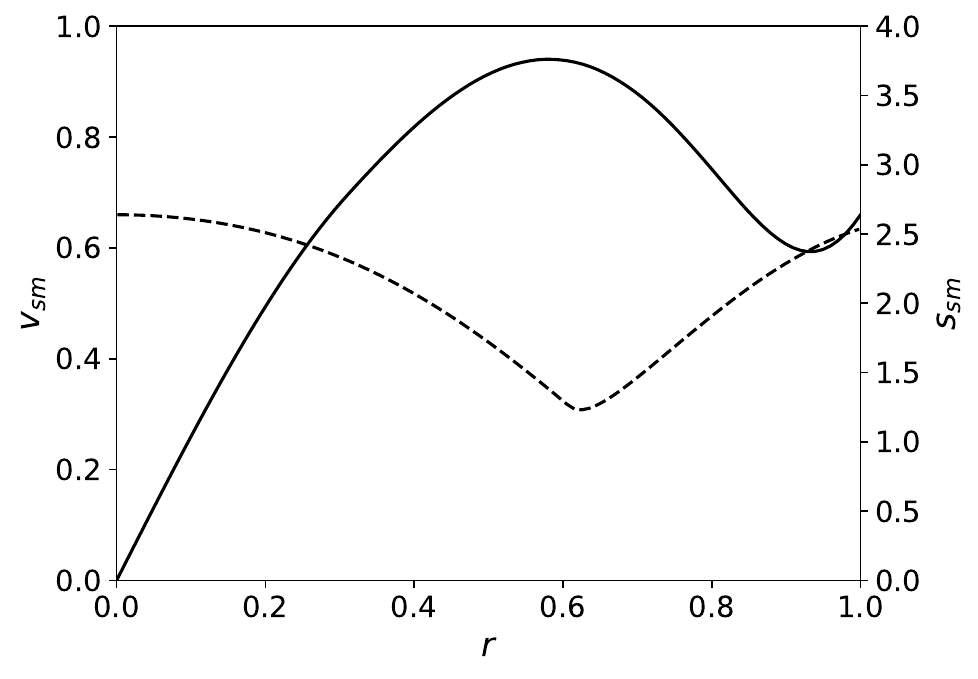}
\caption{Same as Fig.\ 1 for the $s_2t_2$ field $\vv_{D\! J}$ of Dudley 
\& James (1989).}
\end{figure}
The ratio $R_c (\al) /R_{lb} (\al)$ in the Backus- as well as in the Childress-case does not show significant improvement compared to the classical values at $\al = 0$ (see Fig.\ 6).
\begin{figure}
\includegraphics[width=0.8\linewidth]{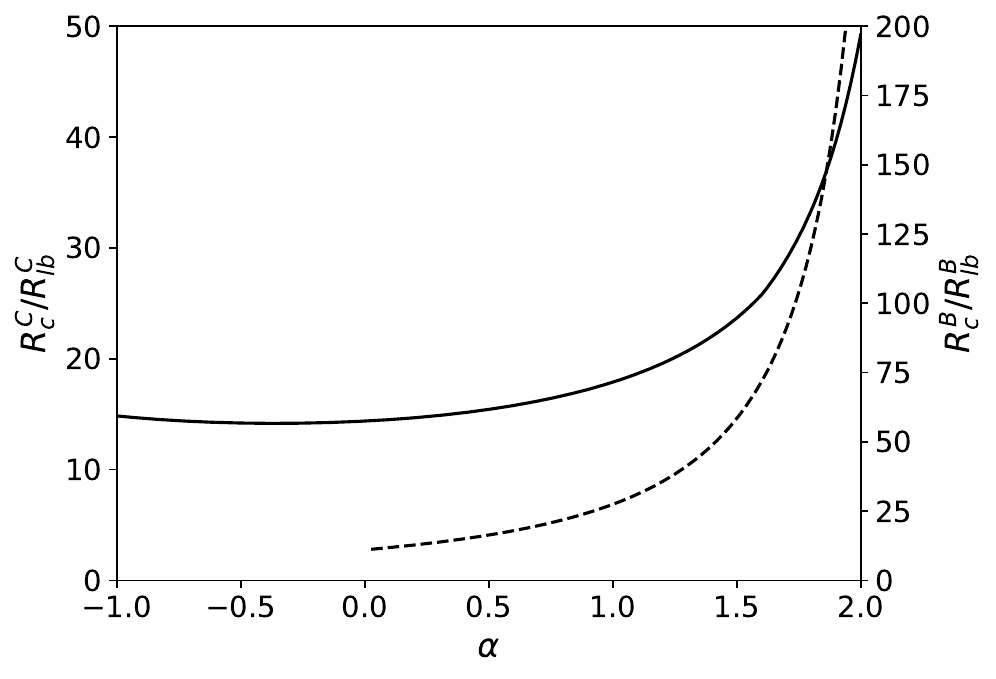}
\caption{Same as Fig.\ 4 for the velocity field $\vv_{D\! J}$.}
\end{figure}
The Backus-ratio seems to be monotonous in the interval $[0,2)$, but, in fact, has a minimum in the range $0.01 < \al < 0.02$. The Childress-type ratio exhibits a shallow minimum at 14.17 for $\al \approx -0.38$, a ratio that differs by less than $1.6\%$ from its classical value.\footnote{Note that $\vv_{D\! J}$ vanishes at the origin so that $R_c^C [\vv_{D\! J} , r^{\al/2}]$ can be finite also for $\al < 0$.}
In conclusion and not quite unexpectedly the power law weight does not lead to notable improvement of bounds on $\vv_{D\! J}$.

The power law weight has been chosen more or less freehand and, of course, one may ask for the best possible (radial) weight $f$ to obtain ratios $R_c^{B/C} [\vv, f]/R_{lb}^{B/C} [\vv, f]$ as small as possible for a given velocity field $\vv$. This question is easily answered considering the following extended Backus-type variational problem: 
\be \label{4.2}
\sup_{0 \neq f \in \cF}\;
\inf_{0 \neq \BB \in \cB}\, 
\frac{\ds\int_{B_1} |\na \times \BB|^2\, \rd v}{\ds \int_{B_1} |\BB|^2\, f^{-1}\, \rd v}
\ee
under the constraint
$$
\sup_{\rr \in B_1}\, \max_{|\xx| = 1} \Big\{\big|\big(S(\rr) \xx\, , \xx\big)\big| f(|\rr|) \Big\} = const. = 1
$$
with $\cF := \{f : (0,1) \ra \R_+ \mbox{ such that the denominator in (\ref{4.2}) is finite }\}$. Defining the spherical maximum $S_{sm}$ of the strain of $\vv$ by 
\be \label{4.3}
S_{sm} (r) := \max_{\rrh \in S_1} \, \max_{|\xx| = 1} \Big\{\big| \big( S(r \rrh) \xx\, ,\xx \big)\big| \Big\} \, ,
\ee
the constraint takes the form
$$
\sup_{0 < r < 1} \{S_{sm}(r)\, f(r) \} = 1 
$$
or
$$
f^{-1} \geq S_{sm} \; \mbox{ on }\, (0,1)\; \mbox{ and  }\; \exists\, r_0 \in [0, 1]\, \mbox{ with }\, f^{-1} (r_0 ) = S_{sm} (r_0)\, .
$$
Thus $f := S_{sm}^{-1}$ everywhere is clearly the optimal choice in (\ref{4.2}) to obtain an as large as possible infimum. For the Backus-type Reynolds number (\ref{2.6}) we then have by construction $R^B = 1$ and the critical ratio is just given by
\be \label{4.4}
R_c^B /R_{lb}^B = \big(R^B_{lb} [S_{sm}^{-1} ]\big)^{-1} .
\ee
Similarly, in the Childress-case the profile $g:=v_{sm}^{-1}$ with
\be \label{4.5}
v_{sm} (r) := \max_{\rrh \in S_1} \, \big\{ |\vv(r \rrh) | \big\} 
\ee
is the optimal weight, and by (\ref{2.2}), (\ref{2.3}), and (\ref{2.8}) we obtain for the critical ratio:
\be \label{4.6}
R_c^C /R_{lb}^C = \big(R^C_{lb} [v_{sm}^{-2} ]\big)^{-1} = \big(R^B_{lb} [v_{sm}^{-2} ]\big)^{-1/2} .
\ee
Using these optimal weights we computed the toroidal minimum $\mu^t$ as well as the first four poloidal minima $\mu_n^p$ for both the most efficient flow $\vv_{me}$ and for $\vv_{D\! J}$. The minimum was always given by $\mu_1^p$, and by (\ref{3.10}) this holds also for large $n$. The computations of $\mu^t$ and $\mu_n^p$ relied again on the ``direct method'' as explained for the power law weight. The result for both ratios, given by (\ref{4.4}) and (\ref{4.6}), and both velocity fields is shown in table 1. In parentheses we have added the classical ratios with constant weight.

$$
\ba{r|cc}
& R_c^B/R_{lb}^B & R_c^C/R_{lb}^C \\[.5ex]
\hline \\[-1ex]
\vv_{me} & 6.4\; (16.6) & 6.05\; (12.6) \\[1ex]
\vv_{D\! J} & 8.33 \; (11.46) & 12.3\; (14.4) 
\ea
$$
$$\mbox{Table 1: Critical ratios with optimal
weight and constant weight (in parentheses).}$$

As expected the larger improvement by a radially varying weight is observed for $\vv_{me}$, which itself exhibits strong radial variation. Note that the improved Backus- and Childress-type ratios are comparable in size for $\vv_{me}$, whereas the unweighted ratios are not. So, applying the optimal weights to $\vv_{me}$ not only lowers the ratios but also makes them more consistent. For $\vv_{D\! J}$ we find a less pronounced improvement of the ratios, which nevertheless in the Backus-case comes close to the ratio for $\vv_{me}$.
\section{Conclusion and outlook}
Lower bounds on those Reynolds numbers that allow dynamo action should give a rough orientation about the minimum size of this quantity. A low ratio $R_c/R_{lb}$ of critical Reynolds number over its lower bound is thus desirable. Classical bounds make use of Reynolds numbers based on pointwise estimates of the velocity field (Childress-type) or its strain (Backus-type). These Reynolds numbers have here been extended to include spherically symmetric weights of two types: a universal weight of power law type and a weight that is optimally adapted to a given velocity field.  
The weighted ratios have been tested by the ``most efficient'' velocity field $\vv_{me}$, which shows strong radial variation, and an efficient velocity field $\vv_{D\! J}$ that does not exhibit such a characteristics.
For $\vv_{me}$ both types of Reynolds numbers with both types of weight show considerable improvement: the largest improvement -- by a factor of $2.6$ -- is obtained by the Backus-type ratio with optimal weight, whereas in absolute numbers the lowest value of about $6.05$ is obtained by the Childress-type ratio with optimal weight. For $\vv_{D\! J}$ almost no improvement could be achieved by the power law weight and only a moderate improvement (factor 1.4 in the Backus-type ratio) by the optimal weight.

Concerning further improvement, especially the observed discrepancy between $\vv_{me}$ and $\vv_{D\! J}$ suggests that a spherically symmetric weight might not be enough. A more general weight that allows lateral variation can, of course, be better adapted to a given velocity field; the lower bound, however, then requires  the solution of a fully three-dimensional minimization problem. 

The use of different norms, especially integral norms, could be another possibility to obtain better ratios $R_c/R_{lb}$. For example, $\vv_{me}$ has been determined by optimization with respect to the enstrophy norm and proper manipulations in the Backus-balance (\ref{2.5a}) would yield, instead of (\ref{2.7}), the ``nonlinear'' variational problem 
\be \label{4.7}
\inf_{0 \neq \BB \in \cB}\, \frac{\ds\int_{B_1} |\na \times \BB|^2\, \rd v}{\ds \bigg(\int_{B_1} |\BB|^4\,  f^{-1}\, \rd v\bigg)^{1/2} } \, ,
\ee
determining a lower bound in this norm. Even with constant weight, problem (\ref{4.7}) constitutes a formidable task since all the separation properties, which allowed the (almost) exact solution of the quadratic case, are now lost.  Luo et al.\ (2020) for their estimate of problem (\ref{4.7}) made use of an optimal Sobolev constant for a scalar version of (\ref{4.7}) (Talenti 1967). The minimizer then exhibits spherical symmetry, a property that in the vectorial problem with divergence-constraint clearly cannot be expected. On the other hand, a fully numerical approach to problem (\ref{4.7}) is possible, however, faces similar imponderables as the search for the most efficient velocity field and is against the spirit of a {\em rigorous} bound.

Finally, let us stress once more that absolute values of critical Reynolds numbers in different norms carry little information. They are vastly different (see, e.g., table 4 in (Chen et al.\ 2018))  and can even be misleading: in terms of the classical Backus-type Reynolds number the critical value for $\vv_{D\! J}$ is lower than that for the most efficient field $\vv_{me}$. Ratios with (as large as possible) lower bounds are a better choice to judge velocity fields for their dynamo onset.
\setcounter{equation}{0}
\renewcommand{\theequation}{A\arabic{equation}}
\section*{Appendix A}
This appendix provides a sequence of test functions $\big(\BB_n\big)_{n \in \N\setminus \{1\}} \subset \cB$ with property 
\be \label{A.1}
\lim_{n \ra \infty}\, \frac{\ds \int_{B_1} \big|\na \ti \BB_n \big|^2\, \rd v}{\ds \int_{B_1} \big|\BB_n \big|^2\, r^{-\al} \, \rd v}\, = 0\, , \qquad \al > 2
\ee
demonstrating thus $R^B_{lb} [r^\al]= 0$ and  $R^C_{lb} [r^\al] = 0$ for 
$\al > 2$.

The test functions are of toroidal type and given by
$$
\BB_n = \na \ti T_n\, \rr\; , \qquad T_n =\left\{
\ba{ccc} 
c_n^\beta\, (n r)^\ga & \mbox{for } & 0 < r \leq 1/n \\[.3em]
r^\bet \, (1-r) & \mbox{for} & 1/n < r < 1
\ea \ri.
$$
with $c_n^\bet := n^{-(\bet +1)} (n-1)$, $n \in \N\setminus \{1\}$, and $\bet$ and $\ga$ yet to be determined. Explicitly we obtain
\be \label{A.2}
\BB_n =  \left\{
\ba{clc} 
c_n^\beta\, (n r)^\ga \sin \theta \, \eee_\ph & \mbox{for } & 0 < r \leq 1/n \\[.4em]
r^\bet \, (1-r) \sin \theta \,\eee_\ph  & \mbox{for} & 1/n < r < 1
\ea \ri.
\ee
and 
\be \label{A.3}
\na \ti \BB_n =  \left\{
\ba{c} 
n\, c_n^\beta\, (n r)^{\ga - 1} \big( 2 \cos \theta \, \eee_r - (\ga + 1) \sin \theta \, \eee_\theta \big) \\[.4em]
r^{\bet -1} \, \big( 2 (1-r) \cos \theta \, \eee_r - (\bet + 1 - (\bet + 2) r ) \sin \theta \,\eee_\theta \big) .
\ea \ri.
\ee
Note that under the condition 
\be \label{A.3a}
\ga > 0
\ee
$\BB_n$ satisfies the boundary conditions of a toroidal magnetic field (with trivial harmonic extension in $\Bh$), $\BB_n $ is weakly differentiable and in fact continuous on $\R^3$. The subsequent calculations prove both integrals in (\ref{A.1}) to be finite under suitable conditions on $\bet$ and $\ga$, thus $\BB_n \in \cB$ (cf.\ definition (\ref{2.3a})). Moreover, in the limit $n \ra \infty$ the denominator in (\ref{A.1}) grows indefinitely whereas the numerator stays bounded, which proves the claim.

In fact, by (\ref{A.2}) one calculates
$$
\ba{l}
\ds \int_{B_1} |\BB_n|^2 \, r^{-\al}\, \rd  v = \bigg( (c_n^\bet)^2\, n^{2 \ga} \int_0^{1/n} r^{2 + 2 \ga - \al}\, \rd r + \int_{1/n}^1 r^{2 + 2 \bet - \al} (1 - r)^2\, \rd r \bigg) 2 \pi  \int_0^\pi \sin^3 \theta\, \rd \theta \\[3ex]
\ds\qquad \geq \bigg( (c_n^\bet)^2\, n^{2 \ga}\, \frac{(1/n)^{3 + 2 \ga - \al}}{3 + 2 \ga - \al} + \frac{1}{4} \int_{1/n}^{1/2} r^{2 + 2 \bet - \al} \, \rd r \bigg) \frac{8\pi}{3} \\[3ex]
\ds \qquad \qquad \geq \frac{8 \pi}{3} \frac{1}{4} \bigg( \frac{(1/2)^{3 + 2 \bet - \al}}{3 + 2 \bet - \al} + \Big( \frac{1}{3 + 2 \ga - \al} - \frac{1}{ 3+2 \bet - \al} \Big) \Big(\frac{1}{n}\Big)^{3 + 2\bet - \al} \bigg) \xrightarrow[n \ra \infty]{} \infty\, ,
\ea
$$
where we made use of the conditions
\be \label{A.4}
3 + 2 \ga - \al > 0\quad \mbox{and}\quad 3 + 2 \beta - \al < 0\, .
\ee
The former condition guarantees integrability at the origin and the latter one the claimed asymptotics for $n \ra \infty$.

Similarly, by (\ref{A.3}) one obtains
$$
\ba{l}
\ds \int_{B_1} |\na \ti \BB_n|^2 \, \rd  v = n^2 (c_n^\bet)^2\, n^{2 (\ga- 1)} \int_0^{1/n} r^{ 2 \ga }\, \rd r \; \frac{8 \pi}{3} \big(2 + (\ga +1)^2\big) + \int_{1/n}^1 r^{2 \bet} \, \rd r  \; \frac{8 \pi}{3} \big(2 + d_\bet\big)\\[3ex]
\ds\qquad \leq \frac{8 \pi}{3} \big(2 + \max \{ (\ga +1)^2, d_\bet \}\big) \bigg( \frac{1}{2 \bet +1} + \Big( \frac{1}{2 \ga + 1} - \frac{1}{2 \bet + 1} \Big) \Big(\frac{1}{n}\Big)^{2 \bet +1} \bigg) \\[3ex]
\ds\qquad \qquad \leq \frac{8 \pi}{3} \big(2 + \max \{ (\ga +1)^2, d_\bet \}\big) \Big( \frac{1}{2 \bet +1} + \frac{1}{2 \ga + 1} \Big) 
= const.
\ea
$$
with $d_\bet := \sup_r (\bet + 1 -(\bet + 2) r)^2$ under the condition 
\be \label{A.5}
2 \bet + 1 \geq 0\, .
\ee
Summarizing conditions (\ref{A.3a}), (\ref{A.4}), and (\ref{A.5}) we have 
$$
0 \leq \bet + 1/2 < \al/2 - 1\; , \quad \ga > \max \{0\, , \al/2 - 3/2\}\, ,
$$
which can be satisfied, e.g., by 
$$
\bet := -1/2 + (\al -2)/4\;, \qquad \ga := \al/2 - 1\, .
$$
\setcounter{equation}{0}
\renewcommand{\theequation}{B\arabic{equation}}
\section*{Appendix B}
This appendix proves inequality (\ref{3.10}), which locates the poloidal minimum $\mu_n^p$ at small $n$. The presentation varies an argument given in (Kaiser \& Tilgner 2018, appendix A). The starting point is a reformulation of the variational expression (\ref{3.8}), viz.,
\be \label{B.1}
\mu_n^p = \inf_{\sh_n \in H_\infty^2} \frac{\ds \int_0^\infty \Big({\shn}'' - \frac{n(n+1)}{r^2}\, \shn\Big)^2 \rd r}{\ds \int_0^1 \Big((\shn')^2 + \frac{n(n +1)}{r^2}\, \shn^2 \Big) f^{-1} \,\rd r} 
\ee
with variational class
$$
\ba{l}
\ds H_\infty^2 := \big\{ \shn : (0,\infty) \ra \R : \shn \mbox{ is twice weakly differentiable, } \\[.4em]
\ds \qquad \qquad \quad \shn (0) = \shn' (0) =0\,\mbox{ and } \int_0^\infty (\shn'')^2\, \rd r < \infty
\bigg\}\, .
\ea
$$
Equation (\ref{B.1}) follows as eq.\ (\ref{3.8}) by orthogonal decomposition starting, however, with the variational expression (\ref{2.4}) instead of (\ref{2.3}). Using integration by parts the numerator in (\ref{B.1}) can now be rewritten as follows:\footnote{Note that boundary terms at $r=0$ and at infinity do not arise for functions of class $H_\infty^2$.}
$$
\ba{l}
\ds \int_0^\infty \Big(\shn'' - \frac{n(n+1)}{r^2}\, \shn\Big)^2 \rd r 
 = \int_0^\infty \Big( (\shn'')^2 - 2\frac{n(n+1)}{r^2}\, \shn \shn'' + \Big(\frac{n(n+1)}{r^2}\Big)^2 \shn^2\Big) \rd r \\[1.2em] 
 \ds \qquad = \int_0^\infty \Big( (\shn'')^2 + 2\frac{n(n+1)}{r^2}\, (\shn')^2 + \frac{n(n+1)}{r^4}\, \big(n (n+1) - 6\big) \shn^2\ \Big) \rd r \\[1.2em]
\ds \qquad \qquad  = \int_0^\infty \Big( (\shn'')^2 + \frac{4}{r^2}\, (\shn')^2 - \frac{8}{r^4}\, \shn^2\, \Big) \rd r \\[1em]
\ds \qquad \qquad \qquad \qquad + \big(n (n+ 1) -2\big)
\int_0^\infty \Big( \frac{2}{r^2}\, (\shn')^2 + \frac{n(n+1)- 4}{r^4}\, \shn^2\, \Big) \rd r\, .
\ea
$$
Writing, similarly, the denominator in the form 
$$
\ba{l}
\ds \int_0^1 \Big( (\shn')^2 + \frac{n(n+1)}{r^2}\, \shn^2\Big) f^{-1}\, \rd r \\[1em]
\ds \qquad \quad = \int_0^1 \Big( (\shn')^2 + \frac{2}{r^2}\, \shn^2\, \Big) f^{-1}\, \rd r + \big(n (n+ 1) -2\big)
\int_0^1 \frac{1}{r^2}\, \shn^2 \, f^{-1}\, \rd r\, ,
\ea
$$
eq.\ (\ref{B.1}) by (\ref{3.4}) can be estimated as follows:
$$
\ba{l}
\ds \mu_n^p = \inf_{\sh_n \in H_\infty^2} \frac{\ds \int_0^\infty \Big({\shn}'' - \frac{n(n+1)}{r^2}\, \shn\Big)^2 \rd r}{\ds \int_0^1 \Big((\shn')^2 + \frac{n(n +1)}{r^2}\, \shn^2 \Big) f^{-1} \,\rd r} \\[2.5em]
\ds \geq  \inf_{\shn \in H^2_\infty}\, \min \Bigg\{
\frac{\ds \int_0^\infty\! \Big(({\shn}'')^2 +  \frac{4}{r^2}\, (\shn')^2 - \frac{8}{r^4}\, \shn^2 \Big) \rd r}{\ds \int_0^1 \Big((\shn')^2 + \frac{2}{r^2}\, \shn^2 \Big) f^{-1} \,\rd r}\, ,
\frac{\ds \int_0^\infty\! \Big(\frac{2}{r^2}\, ({\shn}')^2 + \frac{n(n+1)- 4}{r^4}\, \shn^2 \Big) \rd r}{\ds \int_0^1 \frac{1}{r^2}\, \shn^2 \, f^{-1} \,\rd r}
\Bigg\} \\[2.5em]
\ds \qquad \qquad  \geq \min \Bigg\{\mu_1^p\,  , \,
\inf_{\shn \in H^2_\infty}
\frac{\ds \int_0^\infty\! \Big(\frac{2}{r^2}\, ({\shn}')^2 + \frac{n(n+1)- 4}{r^4}\, \shn^2 \Big) \rd r}{\ds \int_0^1 \frac{1}{r^2}\, \shn^2 \, f^{-1} \,\rd r}
\Bigg\} .
\ea
$$
By the Hardy-type inequalities
\be \label{B.2}
\int_0^\infty \sh^2\, \frac{\rd r}{r^2} \leq 4 \int_0^\infty (\sh')^2 \, \rd r\;  , \qquad \int_0^\infty \sh^2\, \frac{\rd r}{r^4} \leq \frac{4}{5} \int_0^\infty (\sh')^2 \,\frac{\rd r}{r^2}
\ee
and by (\ref{3.9}) we can further estimate:
$$
\frac{\ds \int_0^\infty\! \Big(\frac{2}{r^2}\, ({\shn}')^2 + \frac{n(n+1)- 4}{r^4}\, \shn^2 \Big) \rd r}{\ds \int_0^1 \frac{1}{r^2}\, \shn^2 \, f^{-1} \,\rd r} \geq 
\frac{\ds \Big(n(n+1)- \frac{3}{2} \Big)\int_0^\infty \frac{1}{r^4}\, \shn^2  \rd r}{\ds \frac{1}{m}\int_0^1 \frac{1}{r^4}\, \shn^2 \, \rd r} \\[1em]
\geq m \Big(n(n+1) - \frac{3}{2}\Big)\, ,
$$
which proves inequality (\ref{3.10}).
%
%
%
%
\section*{References}


\vspace{2mm}\noi
Backus, G. E., A class of self-sustaining dissipative spherical dynamos. {\em Ann.\ Phys.} {\bf 4}, 372--447 (1958).


\vspace{2mm} \noi
Chen, L., Herreman, W., Li, K., Livermore, P.W., Luo, J.W. and Jackson, A., The optimal kinematic dynamo driven by steady flows in a sphere. {\em J.\ Fluid Mech.} {\bf 839}, 1--32 (2018).

\vspace{2mm}\noi
Childress, S., {\em Th\'eorie Magn\'etohydrodynamique de l'Effet Dynamo.}, 1969 (Report of D\'e\-parte\-ment M\'ecanique de la Facult\'e des Sciences, Paris).



\vspace{2mm} \noi
Dudley, M.L. and James, R.W., Time-dependent kinematic dynamos with stationary flows. {\em Phil.\ Trans.\ R.\ Soc.\ Lond.\ A}{\bf 425}, 407--429 (1989).


\vspace{2mm} \noi
Holme, R., Optimized axially-symmetric kinematic dynamos. {\em Phys.\ Earth Planet.\ Inter.} {\bf 140}, 3--11 (2003).


\vspace{2mm} \noi
Kaiser, R. and Tilgner, A., On Vainshtein's dynamo conjecture. {\em Proc.\ R.\ Soc.\ Lond.\ A}{\bf 455}, 3139-3162 (1999).

\vspace{2mm} \noi 
Kaiser, R. and Tilgner, A., Optimal energy bounds in spherically symmetric $\al^2$-dynamos. {\em Quarterly of Applied Mathematics }{\bf 76}, 437-461 (2018).



\vspace{2mm} \noi
Love, J. J. and Gubbins, D., Optimized kinematic dynamos. {\em Geophys.\ J.\ Int.} {\bf 124}, 787--800 (1996).

\vspace{2mm} \noi
Luo, J.W., Chen, L., Li, K. and Jackson, A., Optimal kinematic dynamos in a sphere. {\em Proc.\ R.\ Soc.\ Lond.\ A}{\bf 476}, 20190675 (2020).

\vspace{2mm}\noi
Moffatt, H. K., {\em Magnetic Field Generation in Electrically Conducting Fluids} (Cambridge University Press, Cambridge, England 1978).

\vspace{2mm}\noi
Proctor, M. R. E., On Backus' necessary condition for dynamo action in a conducting sphere. {\em Geophys.\ Astrophys.\ Fluid Dynam.} {\bf 9}, 89--93 (1977).

\vspace{2mm}\noi
Proctor, M. R. E., Necessary conditions for the magnetohydrodynamic dynamo. {\em Geophys.\ Astrophys.\ Fluid Dynam.} {\bf 14}, 127--145 (1979).



\vspace{2mm} \noi
Talenti, G., Best constant in Sobolev inequality. {\em Ann.\ Mat.\ Pura Appl.} {\bf 110}, 353--372 (1976).



%
%
%
\end{document}